\documentclass[epj]{svjour}
\usepackage{amsmath}
\usepackage{graphicx}
\usepackage{amssymb}
\usepackage{color}
\usepackage{enumerate}

\usepackage{graphics}
\begin{document}
\title{The Schr$\ddot{o}$dinger-Poisson equations as the large-N limit of the
Newtonian N-body system: applications to the large scale dark
matter dynamics.}

\author{Fabio Briscese}

\institute{Department of Mathematics, Physics and Electrical
Engineering, Northumbria University, Newcastle City Campus,
Ellison Building D 207, NE1 8ST, Newcastle upon Tyne, UK \and
Istituto Nazionale di Alta Matematica Francesco Severi, Gruppo
Nazionale di Fisica Matematica, Citt\`{a} Universitaria, P.le A.
Moro 5, 00185 Rome, Italy.}
\date{Received: date / Revised version: date}
%
\abstract{ In this paper it is argued  how the dynamics of the
classical Newtonian N-body system can be described in terms of the
Schr$\ddot{o}$dinger-Poisson equations in the large $N$ limit.
This result is based on the stochastic quantization introduced by
Nelson, and on the Calogero conjecture. According to the Calogero
conjecture, the emerging effective Planck constant is computed in
terms of the parameters of the N-body system as $\hbar \sim
M^{5/3} G^{1/2} (N/<\rho>)^{1/6}$, where is $G$ the gravitational
constant, $N$ and $M$ are the number and the mass of the bodies,
and $<\rho>$ is their average density. The relevance of this
result in the context of large scale structure formation is
discussed. In particular, this finding gives a further argument in
support of the validity of the Schr$\ddot{o}$dinger method as
numerical double of the N-body simulations of dark matter dynamics
at large cosmological scales.
 }

\maketitle

\section{Introduction}

It is widely accepted tat the formation of   large scale
structures (LSS) in the universe, as  superclusters, sheets and
filaments \cite{LSS}, is shaped by collisionless  Dark Matter (DM)
\cite{DM} (see \cite{DM history} for an historical review of DM).
In the standard $\Lambda$-CDM cosmological model \cite{mukhanov},
DM is assumed to be constituted of unknown particle species that
interact (almost) only gravitationally, and DM is described as a
cold fluid at cosmological scales. In fact, cold  dark matter
(CDM) is in agreement with all cosmological data, including LSS
\cite{LSS}, CMB \cite{planck}, leansing \cite{leansing}, BAO
\cite{bao},  and supernovae \cite{supernovae}. Even though  DM
particles are still elusive, there are currently many DM
candidates, and the search for DM particles is an open issue
\cite{DM particles}.

In this context, the study of the  evolution of CDM becomes
crucial. At large cosmological scales, CDM  is successfully
described  as a pressureless dust fluid; but this assumption fails
at smaller scales, where bound structures form. At small scales,
and for realistic cases where the typical velocities are
non-relativistic,  the Newtonian limit of the Einstein equations
is sufficient to describe the time evolution of massive bodies
within the universe \cite{small vel universe}. Therefore, at cosmological scales CDM  can
be safely  described as a classical N-body system in which the individual particles
represent bounded agglomerates of DM particles that interact only gravitationally.

The first and most natural way to treat this N-body system is to
resort to N-body numerical simulations \cite{n simul,n simul1,n
simul2,n simul3,n simul4}. For instance, the MILLENNIUM simulation
\cite{n simul} was carried out tracing the evolution of $N \sim
10^{10}$ identical particles of mass $M \simeq 10^9 M_{\odot}$,
where $M_{\odot}\simeq 2 \times 10^{30} Kg$ is the solar mass. The
particles in the N-body system represent huge agglomerations of
elementary dark matter particles, and although in the simulation
these particles do all have the same mass, this description is
sufficient to explain how halos with a wide variety of masses and
different abundances are built up from such effective particles.
In fact, the simulation shows the formation of galactic halos made
of hundreds of particles, and of clusters of galaxies made of
millions of particles.

One might argue that the choice of the mass of the particles in
the N-body simulation is arbitrary, and question whether this
choice affects the final results. This does not seem to be the
case \cite{n simul,n simul1,n simul2,n simul3,n simul4}, as far as
the mass $M$ is much smaller than the mass of the objects that we
study; e.g. $M$ should be much smaller than the mass of galaxies
if one wants to study the formation of galactic halos, but it can
be of the order of the (average) mass of galactic halos if one
aims to describe the formation of LSS. Of course, a practical
lower bound on $M$ comes from the fact that decreasing $M$, and
therefore increasing the resolution attained in the N-body
simulation, one increases the computational effort to solve the
N-body dynamics numerically.

Thus, although the choice $M \simeq 10^9 M_{\odot}$ in the N-body
simulations is fit for purpose, it is still arbitrary. However, an
indication on the plausibility of this value comes from scalar
field DM models \cite{matos}. In fact,   a massive and
non-interacting scalar field with lagrangian density $\emph{L} =
\partial_\mu \Phi \partial^\mu \Phi - m_\phi |\Phi|^2$ forms bound
DM halos  \cite{arbey} (the generalization to self-interacting
scalar fields with quartic potential has been studied in
\cite{quartic}). In the case of spherically symmetric scalar field
configurations $\Phi = \exp{[- i m_\phi t]}\sigma(r)/\sqrt{2}$,
the size of the halo is given by $l \sim \sqrt{M_P/\sigma(0)}
\left(\hbar/m_\phi c\right)$. From such relation it is quite
evident that, even if $M_P/\sigma(0)\gg 1$ so that $l$ is much
bigger than the Compton wavelength of DM particles, halos of size
above $\gtrsim 10 kpc$ are formed only for ultralight DM
particles. The typical orbiting velocity in the halo is $v_o/c
\sim \sqrt{\sigma(0)M_P}$, and using $l \sim 10 \, kpc$ and $v_o
\sim 100 \, km/s$ for low luminosity spiral galaxies, one has
$\sigma(0)/M_P \sim 10^{-6}$, while $m \sim 10^{-23} eV$
\cite{arbey}. We stress that this estimate of $m_\phi$ coincides
(in order of magnitude) with that obtained in
\cite{zhang,paredes,shive,urena}. Furthermore, the mass $M$ of the
halo in this simple model is given by $M\sim \sqrt{\sigma(0)
M_P^3/m_\phi^2} \sim  10^9 M_{\odot}$ \cite{arbey},   indeed $m
\sim 10^{-23}$ is an upper bound for the DM particle mass yielding
a lower bound for the masses of halos that can be realized. We
mention that scalar field dark matter might be useful to resolve
potential small scale problems of CDM, see \cite{witten} for an
exhaustive discussion of this issue.

Due to the huge numerical effort to solve the N-body problem in
realistic situations when $N \sim 10^{10}$, it would be desirable
to have a simple analytical model from which it is possible to
extract the most important physical properties of this N-body
system. An alternative is to describe the N-body dynamics
statistically, by means of the phase-space distribution of the
bodies $f(t,x,p)$, where the evolution of $f(t,x,p)$ is given by
the Boltzmann equation. In the case of LSS, N is large and
collisions are suppressed; moreover, the dynamics is only affected
by the Newtonian potential \cite{Gilbert }, so that the Boltzmann
equation reduces to a Vlasov (or collisionless Boltzmann) equation
\cite{andreasson}. Although this model is simple from a conceptual
point of view, there is no general solution of the Vlasov
equation. However, the relevant physical information  can be
extracted from the momenta $M^{(n)}$ of the distribution function.
To do so, one should solve the infinitely coupled hierarchy of
equations for the momenta $M^{(n)}$,  and it turns out that the
only coherent way to neglect higher cumulants is to neglect them
entirely \cite{pueblas};  but in this case one reduces to the dust
model, which we know to be inappropriate  to describe halo
formation, while giving a good description of LSS at larger
scales.

We can assume that, for our purposes, it is sufficient to study
the  evolution of smoothed density and velocity fields
\cite{smothen alternative}. A possibility is to use the so called
Schr$\ddot{o}$dinger method (ScM), which has been proposed as
numerical technique  to describe the dynamics of CDM
\cite{schrodinger,schrodinger1,schrodinger2,schrodinger3,schrodinger4,schrodinger5,schrodinger6,schrodinger7,schrodinger8,schrodinger9,schrodinger10,schrodinger11,zhang,paredes,shive,urena}.

ScM is based on the hypothesis that, in the Newtonian limit, it is
possible to describe the evolution of DM by means of a wave
function $\psi$, such that the DM density is given by $\rho_{DM} =
M |\psi|^2$, where $M$ represents the effective mass of DM
particles. The wave function $\psi$ obeys the coupled
Schr$\ddot{o}$dinger-Poisson equations (SPEs)

\begin{equation}
\begin{array}{ll}\label{SPE equation}
i \hbar \partial_t \psi = - \frac{\hbar^2 }{ 2 a(t) M } \Delta \psi + M V \, \psi \, , \\
\\
\Delta V =  4 \pi G \, \rho  \, ,
\end{array}
\end{equation}
where $V$ is the Newtonian potential, $\rho$ is the energy density
of the universe,  $a(t)$ is a scale factor introduced to take into
account the expansion of the universe, and $\hbar$ is a parameter
representing an effective Planck constant. The Newtonian potential
$V$ is determined through the Poisson equation in (\ref{SPE
equation}). The form of the DM density used to run cosmological
simulations in
\cite{schrodinger,schrodinger1,schrodinger2,schrodinger3,schrodinger4,schrodinger5,schrodinger6,schrodinger7,schrodinger8,schrodinger9,schrodinger10,schrodinger11,zhang,paredes,shive,urena}
is $\rho = (M \, \left\vert\psi\right\vert^2- \rho_{crit} )/a(t)$,
where $\rho_{crit}$ is a parameter representing a comoving
critical density of the universe, although some authors assume
$\rho_{crit}=0$, so that the SPEs reduce to the
Schr$\ddot{o}$dinger-Newton equations \cite{ruffini}. However,
since we are not interested in discussing the explicit form of the
DM density, we will use the generic expression $\rho$.
Furthermore, here we are only focused on the relation between the
ScM and N-body simulations, and this is not related to the
expansion of the universe; thus hereafter we set $a(t) =1$.

The SPEs  can be viewed as the non-relativistic limit of the
Klein-Gordon and Dirac equations, and their theoretical
justification follows from the correspondence principle that
relates classical and quantum mechanical phase-space-distribution
functions in the semiclassical limit  \cite{takahashi}. In this
case $\hbar$ coincides with the Planck constant, and $M$ is the
mass of the elementary DM particles. Numerical solutions of the
SPEs point towards  ultralight DM particles of mass $M \sim
10^{-23} \, eV$  \cite{zhang,paredes,shive}.

Alternatively, the ScM has been introduced as a numerical double
of the N-body description of DM at large scales; see e.g.
\cite{schrodinger,schrodinger1,schrodinger11} and references
therein. In this case, the effective particles in the N-body
system are again huge agglomerates of DM elementary particles, and
therefore $M$ is huge in comparison with the mass of elementary DM
particles. Moreover, the constant $\hbar$ in (\ref{SPE equation})
does not coincide with the Planck constant, nor is fixed by the
N-body problem, but it is merely a free parameter that can be
chosen at will.  Furthermore, due to the \textit{correspondence
principle} \cite{takahashi},  $\hbar$ determines the phase-space
resolution in the ScM. To ensure the match with N-body simulations
one must require $\hbar/M \sim 10^{-4} Mpc \cdot c$, so that
$\hbar$ is huge in comparison with the true Planck constant.

It is worth to emphasize the difficulty in reconciliating such
huge values of $M$ and $\hbar$ with the derivation of the SPEs
from the fundamental quantum mechanical evolution of DM particles.
In fact, even though one might assume that $M$ represents the mass
of huge agglomerates of DM particles, one encounters the
insurmountable problem of explaining the extremely large value of
$\hbar$. In fact, since the Planck constant is fundamental, its
value should not be affected by the Newtonian and semiclassical
limits.

In this paper we discuss the relation between the ScM and the
N-body description of  DM used in numerical simulations \cite{n
simul,n simul1,n simul2,n simul3,n simul4}, and we argue how the
SPEs can be obtained as the large $N$ limit of the Newtonian
N-body system. This argument is valid beyond the context of LSS
formation, and it implies that any Newtonian N-body system of
identical bodies can be described by means of the SPEs, and that
makes this finding of wide interest. For completeness, we mention
that the correspondence between the ScM method and the Vlasov
equation has been extensively studied, and we remand the
interested reader to the existing literature; see
\cite{schrodinger11} and references therein.

Our staring point is the Newtonian N-body system of DM
agglomerates considered in \cite{n simul,n simul1,n simul2,n
simul3,n simul4}. We show that the dynamics of this system
satisfies the hypothesis of the so called Nelson stochastic
quantization \cite{nelson} in the large N limit. That implies that
the evolution of the system can be described statistically, by
means of the Schr$\ddot{o}$dinger equation. The stochastic
background responsible for the Nelson quantization is given, as in
the Calogero conjecture, by the gravitational interaction between
the N bodies, and its stochastic character  is due to the chaotic
behavior of the N-body dynamics. What is more, the Calogero
conjecture also allows to estimate the order of magnitude of the
effective Planck constant.

To begin, let us discuss briefly the hypothesis of the Nelson
stochastic quantization \cite{cetto}, and let us consider a
particle of mass $M$ which moves according to the Newton laws of
motion. The further assumption is that this particle constantly
undergoes a Brownian motion with no friction, and with a diffusion
coefficient $\hbar/M$ inversely proportional to its mass $M$.
Therefore, the trajectory of this particle will be given by

\begin{equation}\label{nelson newton}
M \, \ddot{\vec{x}} = - \vec{\nabla} \phi + \vec{B}(t)
\end{equation}
where $- \vec{\nabla} \phi$ represents all the  conservative
forces, and $\vec{B}(t)$ is a random variable with zero mean,
representing a small random noise. Nelson has shown \cite{nelson}
that, under these hypotheses, the motion of the particle can be
described by means of  a stochastic process, and the probability
distribution $f(x)$ of the particle can be expressed as $f(x) =
\left\vert \psi \right\vert^2$, in terms of a wave function $\psi$
satisfying the Schr$\ddot o$edinger equation

\begin{equation}
\begin{array}{ll}\label{schrodinger}
i \hbar \partial_t \psi = - \frac{\hbar^2 }{ 2 M  } \Delta \psi + \phi \, \psi \, . \\
\end{array}
\end{equation}

Therefore, in this picture the quantum behavior of the dynamics of
the particle is not a fundamental property of the nature, but it
is induced by the random field $B(t)$. It is necessary to note
that the Nelson quantization only implies the emergence of the
Schr$\ddot o$edinger equation (\ref{schrodinger}), which of
course, does not encompass all the features of quantum mechanics.
For instance, all the properties related to the measurement
processes in quantum mechanics, e.g. entanglement, have not been
derived in the context of Nelson quantization. However, we are not
interested in discussing the validity of the Nelson quantization
as a real theory of quantum world, and we refer the reader
interested in this problem to the literature (see \cite{cetto} and
references therein); but we want to exploit the result of Nelson
in the context of the N-body dynamics.

At this point, one can ask the question of the nature of the
random field $B(t)$ responsible for the emergent quantization. One
of the most studied possibilities is that $B(t)$ is  the random
zero-point radiation field  of the electromagnetic field
\cite{cetto}. Another possibility, conjectured by Calogero
\cite{calogero conjecture}, is that the random noise $B(t)$ is the
resultant of the gravitational interaction of the particle with
all the other particles of the universe. In fact, apart from the
interaction with neighboring bodies which is not small and must be
included in the potential $\phi$,  the gravitational interaction
with far bodies behaves as a small background noise, and its
random behavior comes from the fact that the classical dynamics of
a N-body system is chaotic (see for instance \cite{strogatz} for a
review of classical chaos).   In the context of cosmological
simulations, it has been shown that  chaos appears at scales
smaller than a critical transition scale $\sim 3.5 Mpc/h$, where
$h$ is the dimensionless Hubble parameter, while the dynamics
appears to be nonsensitive to initial conditions (thus
non-chaotic) at larger scales; see \cite{chaos cosmo} for more
details.

Therefore, any particle of the N-body system  experiences a
stochastic gravitational acceleration due to the rest of the
system. Exploiting this idea, Calogero has shown that the order of
magnitude of the induced Planck constant is \cite{calogero
conjecture}

\begin{equation}\label{hbar calogero}
\hbar \sim M^{5/3} G^{1/2} (N/<\rho>)^{1/6} \, ,
\end{equation}
where $N$ and $M$ are the number and the mass of the bodies,
$<\rho>$ is the average density of the system, and $G$ the
gravitational constant.

Let us briefly describe how (\ref{hbar calogero}) can be obtained
on the basis of semiquantitative arguments. The relevant
quantities in our analysis are the dimensional parameters $G$,
$M$, and $<\rho>$; plus $N$, which of course is dimensionless.
From these quantities we can define the unit of time $T$ as

\begin{equation}\label{T}
T \sim (G <\rho>)^{-1/2} \, .
\end{equation}
We want to estimate the characteristic time $\tau$ of the
stochastic acceleration that each particle of the system undergoes
due to all the other particles. Since the N-body dynamics is
chaotic, it is plausible that the characteristic frequency of this
motion $\nu \sim 2\pi/\tau$ should be a growing function of $N$,
and since the background gravitational noise is due to a
collective stochastic effect, it is also plausible to assume that
$\nu$ is proportional to the square root of $N$, so that

\begin{equation}\label{tau}
\tau \sim N^{ -1/2 } \, T \,  .
\end{equation}
The "quantum" of action (that is, the characteristic action)
associated with the stochastic gravitational noise is obtained
multiplying $\tau$ by the gravitational energy per particle
$\epsilon$, which is estimated as
\begin{equation}\label{epsilon}
\epsilon \sim G (N M)^2 R^{-1} N^{-1} \sim G M^{5/3} N^{2/3}
<\rho>^{-1/3} \, ,
\end{equation}
where the length $R \equiv (N M/<\rho>)^{1/3}$ represents the
average linear size of the N-body system. Therefore, the effective
Planck constant is obtained as $\hbar \equiv \epsilon \, \tau$,
which finally gives Eq. (\ref{hbar calogero}). At that point we
should emphasize that this argument is not a pure dimensional
analysis since, even though the exponents of the dimensional
quantities in (\ref{hbar calogero}) are fixed by their dimensions,
the dependence on the dimensionless quantity $N$ is fixed by the
assumption made in (\ref{tau}), which  plays a fundamental role in
the derivation of (\ref{hbar calogero}). We stress that
(\ref{tau}) can be justified in a more rigorous way, and we refer
the reader to \cite{calogero conjecture}, where this relation has
been derived through a more detailed analysis of the properties of
the Newtonian N-body system.


Let us come back to our gravitational N-body system. Using the
Calogero conjecture, we have argued that, due to the classical
gravitational interaction with all the other particles, any
particle in the system undergoes a stochastic gravitational noise
which plays the role of the stochastic random noise $B(t)$ in
(\ref{nelson newton}). Thus, the dynamics of each particle of the
system is given in terms of a wavefunction $\psi$ solution of
(\ref{schrodinger}), where $\hbar$ is given by (\ref{hbar
calogero}). At that point, we can express the wavefunction of the
entire system using the Hartree-Fock approximation, so that the
number density of the N-body system will be $n(x) = N \left\vert
\psi \right\vert^2$. This is the analog of the derivation of the
famous Gross-Pitaevskii equation \cite{dalfovo} for a
Bose-Einstein condensate  by means of the Hartree-Fock
approximation, see \cite{ps} for a detailed analysis of the
quantum many-body system of bosons. Finally, the potential in
(\ref{schrodinger}) is $\phi = M \, V$, where $V$ is the
gravitational potential solution of the Poisson equation $\Delta V
= 4 \pi G \rho$, where $\rho = M \, n(x)$.

Of course, this analysis is accurate only for $N$ large, and
therefore we conclude the dynamics of the N-body system is well
described by the SPEs (\ref{SPE equation}) in the large N limit.
Finally, we stress that the resolution of the SPEs in the
phase-space is fixed by $\hbar$, which in turn is fixed by the
corresponding N-body problem.  However, for given values of
$<\rho>$ and $N$, this resolution is improved decreasing $M$,
which is also true for the corresponding N-body system.

We can now exploit this result in the context of the LSS
formation. In fact, if we come back to the description of the CDM
dynamics at large scales as a Newtonian system of $N \sim 10^{10}$
bodies (which in our case represent huge aggregations of DM
particles) of mass $M$, which interact only gravitationally, we
immediately realize that the hypothesis of the Nelson quantization
are satisfied, as in the Calogero conjecture. In this picture, the
gravitational interaction produces the background random field
$\vec{B}(t)$, which in turn induces the Nelson quantization, and
this fact justifies the quantum mechanical treatment of the system
by means of the SPEs (\ref{SPE equation}).

The advantage of this deduction of SPEs from the N-body dynamics
is that we can use the Calogero result (\ref{hbar calogero})  to
estimate the order of magnitude of $\hbar$ in terms of the
parameters of the N-body problem, so that $\hbar$ is no longer a
free parameter.  In a virialized system of size $L$ with velocity
dispersion $\sigma$, the resolution in phase-space in a Schr$\ddot
o$edinger code is given by $\Delta x \, \Delta v \sim \sigma L
/N_G$, where $N_G = L/d$ is the number of grid points and $d$ is
the grid spacing in the simulation
\cite{schrodinger,schrodinger1}. This estimate must be compared
with the value $\Delta x \, \Delta v \sim \hbar/M \sim  M^{2/3}
G^{1/2} (N/<\rho>)^{1/6}$ obtained from (\ref{hbar calogero}). For
instance, in the case of the Millennium simulation \cite{n simul},
where the N-body problem is solved for $N \simeq 10^{10}$
particles of mass $M \simeq 10^9 M_{\odot}$, using $<\rho> \simeq
3 H_0^2/8\pi G \simeq 4 \times 10^{-26} Kg/m^3$, where $H_0 \simeq
h^{-1}\times 100 \times km/ s Mpc$, with $h \simeq 0.73$, is the
Hubble constant, one has $\hbar \simeq 2 \times 10^{66} Kg \,
m^2/s$. This corresponds to a value $\hbar/M \simeq 10^{-4} Mpc
\cdot c$ in the range of values used in numerical simulations,
e.g. $\hbar/M \sim 10^{-4} Mpc \cdot c$ in
\cite{schrodinger11,paredes} or $\hbar/M \sim 10^{-6} Mpc \cdot c$
in \cite{zhang,shive}.

In conclusion, in this paper it has been shown  that the dynamics
of the classical Newtonian N-body system is well described in
terms of the SPEs in the large $N$ limit. This is due to the
stochastic quantization of the N-body system induced by the random
gravitational background produced by the N bodies, as in the
Calogero conjecture. Moreover, the emerging effective Planck
constant in the SPEs can be computed by means of (\ref{hbar
calogero}) in terms of the parameters of the corresponding N-body
system.

When applied to LSS formation, this finding gives a further
argument in support of the validity of the Schr$\ddot{o}$dinger
method as numerical double of the  N-body simulations of DM
dynamics at large cosmological scales \cite{n simul,n simul1,n
simul2,n simul3,n simul4}, and offers a natural justification for
the huge value of $\hbar$ often used in numerical solutions of
SPEs. These results are particularly remarkable, since this
derivation of SPEs in the context of the Schr$\ddot{o}$dinger
method is the first practical application of the Nelson
quantization and of the Calogero conjecture to a realistic
physical problem.

During the proofreading of this manuscript, the author has noticed
a paper \cite{chavanis}, where it has been presented a generalized
Schr$\ddot o$dinger equation derived from the theory of scale
relativity, and its application to the problem of dark matter
halos formation has been discussed. Due to the links to the
results presented in this manuscript, such paper has been included
in the literature

\textbf{Acknowledgments:} the author is very grateful to F.
Calogero for useful discussions on the Calogero conjecture, and to
G. Rigopopulos and I. Moss for useful discussions on the draft
version of this paper.

\end{document}